\documentclass[twocolumn,aps,superscriptaddress,showpacs,floatfix]{revtex4}
\usepackage{amssymb}
\usepackage{amsmath}
\usepackage{graphicx}
\usepackage[normalem]{ulem}
\usepackage[dvips]{color}

\usepackage{CJK}                     
\usepackage{bm}                      
\usepackage{mathptmx}                
\usepackage{dcolumn}                 

\begin{document}

\title{Short-range tensor interaction and high-density nuclear symmetry energy}

\author{Ang Li}
\affiliation{Department of Physics and Astronomy, Texas A$\&$M
University-Commerce, Commerce, Texas 75429-3011,
USA}\affiliation{Department of Physics and Institute of
Theoretical Physics and Astrophysics, Xiamen University, Xiamen
361005, China}
\author{Bao-An Li\footnote{Corresponding author, Bao-An\_Li$@$Tamu-Commerce.edu}}
\affiliation{Department of Physics and Astronomy, Texas A$\&$M
University-Commerce, Commerce, Texas 75429-3011, USA}

\date{\today}

\begin{abstract}
Effects of the short-range tensor interaction on the
density-dependence of nuclear symmetry energy are examined by
applying an approximate expression for the second-order tensor
contribution to the symmetry energy derived earlier by G.E. Brown
and R. Machleidt. It is found that the uncertainty in the
short-range tensor force leads directly to a divergent
high-density behavior of the nuclear symmetry energy.
\end{abstract}

\maketitle


The density dependence of nuclear symmetry energy $E_{\rm sym}(\rho)$ encodes the energy
related to neutron-proton asymmetry in the Equation of State (EOS) of
nuclear matter. While the $E_{\rm sym}(\rho)$ is very important for both nuclear physics 
and astrophysics~\cite{Li98,Li01,Dan00,Bar05,Li08,Sum94,Lat01,Ste05,Xu10,Che05,Tsa09,Cen09,Xia09,Xu10b},
it is still rather uncertain especially at supra-saturation densities. Besides promising constraints being extracted from astrophysical 
observations~\cite{Ste10}, significant progress has been made recently in constraining 
the $E_{\rm sym}(\rho)$ around and below the nuclear matter saturation density $\rho_0$ using 
experiments in terrestrial laboratories~\cite{Lynch11}. Looking forward, it is very exciting 
to note that dedicated experiments are currently underway or being planned at several advanced radioactive
ion beam facilities at CSR/China~\cite{Xiao}, FRIB/USA~\cite{SEP},
GSI/Germany~\cite{Roy}, RIKEN/Japan~\cite{Riken} and
KoRIA/Korea~\cite{Hong} to pin down the high-density behavior of
the $E_{\rm sym}(\rho)$. While essentially all existing many-body
theories have been used to predict the $E_{\rm sym}(\rho)$, the
results diverge quite widely especially at supra-saturation
densities, see, e.g., ref. \cite{Li08} for a recent review. 
Thus, it is necessary to identify fundamental reasons for the uncertain high-density
behavior of the $E_{\rm sym}(\rho)$. Generally speaking, besides the
different techniques often used in treating nuclear many-body problems in various theories,
our poor knowledge about the isospin dependence of the in-medium
nuclear strong interaction is at least partially responsible for the uncertain $E_{\rm sym}(\rho)$. In
fact, it has been recognized that the spin-isospin dependence of
the three-body force, see, e.g., ref. \cite{Xu10,Zuo02}, the
isospin dependence of short-range nucleon-nucleon correlation
functions, see, e.g., ref.~\cite{Xu11}, and the short-range tensor
force, see, e.g., ref.~\cite{pan72} all play some significant roles in
determining the high-density behavior of the $E_{\rm sym}(\rho)$. In particular, it is easy to understand qualitatively why
the nuclear tensor interaction is important in determining the $E_{\rm sym}(\rho)$. 
Within the parabolic approximation of the EOS of isospin asymmetric nuclear matter, see, e.g.,
ref.~\cite{bom91}, the $E_{\rm sym}(\rho)$ can be written as the
difference between the nucleon specific energy in pure neutron matter
(PNM) and symmetric nuclear matter (SNM), i.e., $E_{\rm sym}(\rho)
= E_{\rm PNM}(\rho) - E_{\rm SNM}(\rho)$. It is well known that in
the isospin-singlet $T=0$ nucleon-nucleon interaction channel relevant for
calculating the EOS of SNM, a significant tensor component is required to understand properties of 
deuteron and neutron-proton scattering data, see, e.g., refs.~\cite{mac89,mac01}. Moreover, it has been found
consistently in microscopic many-body calculations that the $T=0$
channel dominates the potential contribution to the symmetry
energy~\cite{bom91,die03}. In this note, using several typical
and widely used tensor forces that are the same at long-range but
have characteristically different short-range behaviors, we
examine effects of the short-range tensor force on the $E_{\rm sym}(\rho)$. 
Applying an approximate expression for the second-order tensor contribution to
the symmetry energy derived earlier by G.E. Brown and R.
Machleidt~\cite{mac94}, we find that the uncertainty in the
short-range tensor force contributes significantly to the divergence
of the $E_{\rm sym}(\rho)$ at supra-saturation densities. 

In the best-studied phenomenology of nuclear forces, i.e., the
one-boson-exchange model, the tensor interaction results from 
exchanges of the isovector $\pi$ and $\rho$ mesons. For instance,
the tensor part of the one-pion exchange potential (OPEP) can be
written in configuration space as~\cite{mac89}
\begin{eqnarray}
 V_{t\pi}&=& -\frac{f_{\pi}^2}{4\pi}m_{\pi}(\tau_1\cdot\tau_2)S_{12}
\nonumber \\~~~~~&&
[\frac{1}{(m_{\pi}r)^3}+\frac{1}{(m_{\pi}r)^2}+\frac{1}{3m_{\pi}r}]\exp(-m_{\pi}r)
\label{pi}
\end{eqnarray}
where $r$ is the inter-particle distance and
$S_{12}=3\frac{(\sigma_1 \cdot r)(\sigma_1 \cdot
r)}{r^2}-(\sigma_2\cdot\sigma_2)$ is the tensor operator. The
$\rho$-exchange tensor interaction $V_{t\rho}$ has the same
functional form as the OPEP, but with the $m_{\pi}$ replaced
everywhere by $m_\rho$, and the $f_{\pi}^2$ by $-f_{\rho}^2$. The
magnitudes of both the $\pi$ and $\rho$ contributions grow quickly
with decreasing $r$. A proper cancelation of the opposite
contributions from the $\pi$ and $\rho$ exchanges is supposed to give a realistic strength for the nuclear tensor
force. However, since the tensor coupling is not well determined
consistently from deuteron properties and/or nucleon-nucleon
scattering data, the tensor interaction is by far the most
uncertain part of the nucleon-nucleon interaction~\cite{mac01}. 
Moreover, it is also possible that the in-medium $\rho$ meson mass $m_{\rho}$ is different from its
free-space value~\cite{BR}. A density-dependent in-medium $m_{\rho}$ will lead to very
different short-range tensor force~\cite{BR90} and affects the symmetry energy at high densities~\cite{Xu10,Don,Rho11}. 
While there is no community-wide consensus on whether the $m_{\rho}$ changes or not in the dense medium, 
it is a possible origin for the uncertain short-range tensor force. 
In addition, due to both the physical and mathematical differences in
construction~\cite{mac01}, various realistic
nuclear potentials usually have widely different tensor components
at short range ($r\leq$0.8 fm). For example, in the Paris
potential~\cite{Paris}, it is just described simply by a constant
soft core. The Argonne V18 (Av18) uses local functions of
Woods-Saxon type~\cite{av18}, while Reid93 applies local Yukawas
of multiples of the pion mass ~\cite{Reid93}. While it is
promising that new experiments, such as, (p,d) reactions induced
by high energy protons~\cite{tan10} or two nucleon knockout
reactions induced by high energy electrons~\cite{Jlab1,Jlab2}, may
allow us to better constrain the short-range tensor force in the
near future, currently the short-range behavior of the tensor
force is still very uncertain. 

It is easy to see from Eq.~(\ref{pi}) that the expectation value
of the tensor force $<V_t>$ is zero. Thus, the first-order tensor
force does not contribute to the symmetry energy unless one
assumes that all isosinglet neutron-proton pairs behave as bound
deuterons with $S_{12}=2$~\cite{Xu10}. In fact, it is the
second-order tensor contribution that is important for the binding
energy of nuclear matter~\cite{kuo65,bro81} and thus also for the
symmetry energy~\cite{mac94}. Using a second-order effective
tensor interaction obtained first by Kuo and Brown~\cite{kuo65},
see. e.g., ref.~\cite{sob} for a review, Brown and Machleidt found
that the tensor contribution to the symmetry energy is
approximately
\begin{equation}\label{Mac}
<V_{sym}> = \frac{12}{e_{\rm eff}}<V_t^2(r)>
\end{equation}
where $e_{\rm eff}\approx 200$ MeV and $V_t(r)$ is the radial part of the tensor force~\cite{mac94}. While this
approximate expression may lead to symmetry energies
systematically different from predictions of advanced microscopic
many-body theories using various interactions, it is handy to
evaluate effects of the different short-range tensor forces within
the same simple and analytical approach. Of course, it is
necessary and also interesting to evaluate the accuracy of
Eq.~(\ref{Mac}) with respect to microscopic many-body calculations
using the same interaction.

To apply Eq.~(\ref{Mac}) we evaluate the expectation value of
$V_{\rm sym}$ using the free single-particle wave
function $(V^{-1}e^{i \textbf{k}\cdot \textbf{r}})\eta_{
\lambda}\zeta_{\tau}$, where $\eta_{\lambda=\uparrow/\downarrow}$
and $\zeta_{\tau=p/n}$ is the spin and isospin wave function,
respectively. The direct and exchange matrixes are, respectively,
\begin{eqnarray}
&& \langle~ \textbf{k} \lambda \tau \textbf{k}' \lambda'
\tau'|V_{{\rm sym}}|\textbf{k} \lambda \tau \textbf{k}' \lambda'
\tau'\rangle \nonumber
\\ &=& \frac{1}{V^2} \int d^3r\int d^3r' e^{-i
\textbf{k}\cdot \textbf{r}} e^{-i \textbf{k}'\cdot \textbf{r}'}
\eta_{\lambda}^{\dagger} (1) \eta_{\lambda'}^{\dagger} (2)
\zeta_{\tau}^{\dagger}(1) \zeta_{\tau'}^{\dagger}(2) \nonumber
\\ &&\times
V_{{\rm sym}}(1,2) e^{i\textbf{k}\cdot \textbf{r}} e^{i
\textbf{k}'\cdot \textbf{r}'} \eta_{\lambda} (1) \eta_{\lambda'} (2)
\zeta_{\tau}(1) \zeta_{\tau'}(2)\nonumber
\\ &=& \frac{1}{V} \int
V_{{\rm sym}}(\textbf{r})d^3 r \\
{\rm and} \nonumber \\
&& \langle \textbf{k} \lambda \tau \textbf{k}' \lambda'
\tau'|V_{{\rm sym}}|\textbf{k}' \lambda' \tau' \textbf{k} \lambda
\tau \rangle \nonumber
\\ &=& \frac{1}{V^2} \int d^3r\int d^3r' e^{-i
\textbf{k}\cdot \textbf{r}} e^{-i \textbf{k}'\cdot \textbf{r}'}
\eta_{\lambda}^{\dagger} (1) \eta_{\lambda'}^{\dagger} (2)
\zeta_{\tau}^{\dagger}(1) \zeta_{\tau'}^{\dagger}(2) \nonumber
\\ &&\times
V_{{\rm sym}}(1,2) e^{i \textbf{k}'\cdot \textbf{r}} e^{i
\textbf{k}\cdot \textbf{r}'} \eta_{\lambda'} (1) \eta_{\lambda}
(2)\zeta_{\tau'}(1) \zeta_{\tau}(2)\nonumber
\\ &=& \frac{1}{V} \delta_{\lambda\lambda'}\delta_{\tau\tau'} \int
\exp[-i(\textbf{k}-\textbf{k}')\cdot \textbf{r}] V_{{\rm
sym}}(\textbf{r})~d^3 r.
\end{eqnarray}
The expectation value of $V_{{\rm sym}}$ in the $S = 1, T = 0$
channel is thus
\begin{eqnarray}
&& <V_{{\rm sym}}> \nonumber
\\ &&=  \frac{1}{16} \frac{1}{2} \sum_{\textbf{k} \lambda \tau}
\sum_{\textbf{k}' \lambda' \tau'} [\langle \textbf{k} \lambda \tau
\textbf{k}' \lambda' \tau'|V_{{\rm sym}}|\textbf{k} \lambda \tau
\textbf{k}' \lambda' \tau'\rangle   \nonumber
\\ &&~~~- \langle \textbf{k} \lambda \tau
\textbf{k}' \lambda' \tau'|V_{{\rm sym}}|\textbf{k}' \lambda' \tau'
\textbf{k} \lambda \tau \rangle] \nonumber
\\&& = \frac{1}{32}\sum_{\textbf{k} \lambda \tau}
\sum_{\textbf{k}' \lambda' \tau'} \frac{1}{V} \{\int V_{{\rm
sym}}(\textbf{r})d^3 r \nonumber
\\ &&~~~-\delta_{\tau\tau'}\delta_{\lambda\lambda'}\int
\exp[-i(\textbf{k}-\textbf{k}')\cdot \textbf{r}] V_{{\rm
sym}}(\textbf{r})~d^3 r\}
 \nonumber
\\ &&=\frac{V}{2}\frac{1}{(2\pi)^6}\int^{k_F}d^3k\int^{k_F}d^3k'
\{\int V_{{\rm sym}}(\textbf{r})d^3 r  \nonumber
\\ &&~~~-\frac{1}{4}\int
\exp[-i(\textbf{k}-\textbf{k}')\cdot \textbf{r}] V_{{\rm
sym}}(\textbf{r})~d^3 r\}.
 \label{eq:vsymm}
\end{eqnarray}

Noticing that the momentum integral
\begin{eqnarray}
\int^{k_F}d^3k e^{i \textbf{k}\cdot \textbf{r}} &=& 4\pi
\int_0^{k_F} k^2 j_0(kr)dk \nonumber
\\&=& \frac{4\pi k_F^3}{3} \frac{3j_1(k_Fr)}{k_Fr}
 \label{eq:int}
\end{eqnarray}
and the particle number density
$\frac{A}{V}=\frac{2}{3\pi^2}k_F^3$, we can write the tensor
contribution to the symmetry energy as
\begin{eqnarray}\label{Epot}
\frac{<V_{{\rm sym}}>}{A} &=& \frac{12}{e_{\rm
eff}}\cdot\frac{k_F^3}{12\pi^2} \{\frac{1}{4}\int V_{t}^2(r)d^3 r
\nonumber
\\&&- \frac{1}{16}
\int[\frac{3j_1(k_Fr)}{k_Fr}]^2V_{t}^2(r)d^3 r\}.
\end{eqnarray}
For large $k_F$, the second integral in the above equation
approaches zero, the first term is thus expected to dominate at
high densities, leading to an almost linear density dependence.

\begin{figure}[htb]
\centering
\includegraphics[width=8cm,height=6cm]{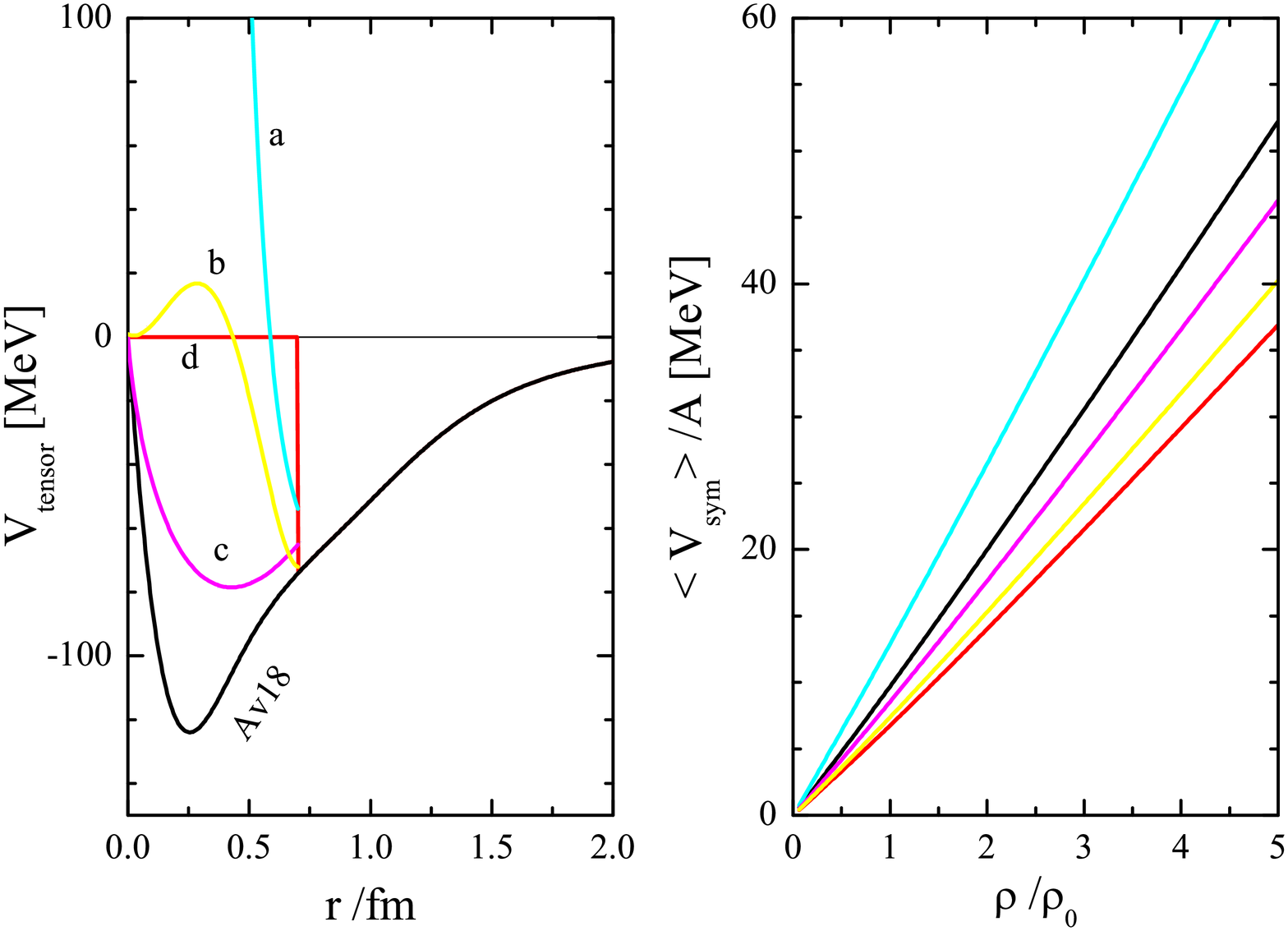}
\caption{(Color online) Left panel: radial parts of the tensor interactions having different short-range behaviors but the same long-range ($r > 0.7$fm) part as the Av18, 
Right panel: potential part of the symmetry energy with the
different short-range tensor interactions.} \label{f1}
\end{figure}

To access quantitatively effects of the short-range tensor force
on the density dependence of nuclear symmetry energy, we adopt
here several tensor forces used by Otsuka et al. in their recent
studies of nuclear structures~\cite{ots05}. The considered tensor
forces, including the standard $\pi+\rho$ exchange (labelled as
$a$), the G-Matrix (GM)~\cite{ots05} (labelled as $b$),
M3Y~\cite{m3y}(labelled as $c$) and the Av18~\cite{av18} (labelled
as Av18), as shown in the left panel of Fig.~\ref{f1}, behave
rather differently at short distance, but merge to the same Av18
tensor force at longer range. In addition, we add a case ($d$)
where the tensor force vanishes for $r\leq 0.7$ fm. The $\pi+\rho$
exchange interaction is fixed by the standard meson-nucleon
coupling constants with a strong $\rho$ coupling~\cite{sob}, and
we use a short-range cut-off at $r=0.4$ fm, i.e.,  $V(r<0.4{\rm
fm})= V(r=0.4{\rm fm})$. As emphasized by Otsuka et al.~\cite{ots05},
the short-range behavior of the tensor force has no effect on 
nuclear structures. However, as we shall show in the following, it affects significantly the
$E_{\rm sym}(\rho)$ especially at supra-saturation densities.

\begin{figure}[htb]
\centering
\includegraphics[width=8cm,height=6cm]{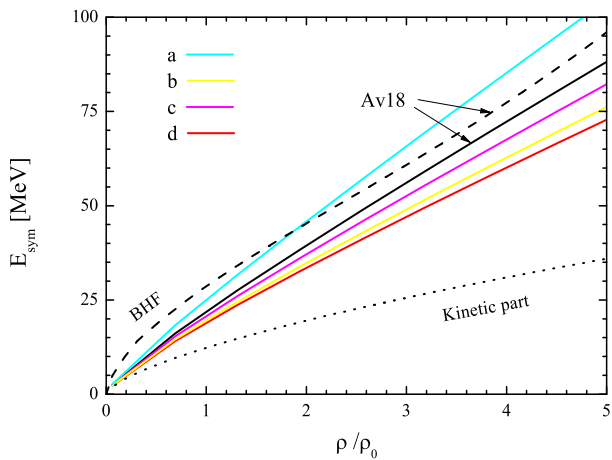}
\caption{(Color online) Symmetry energies using various
short-range tensor interactions in Eq.~(\protect{\ref{Mac}}) in
comparison with the Brueckner-Hartree-Fock prediction using the
Av18 potential.} \label{f2}
\end{figure}
Shown in the right panel of Fig.~\ref{f1} are the potential parts of the symmetry energies 
due to the tensor forces considered according to Eq.~(\ref{Epot}). 
As expected, they tend to grow linearly with increasing density. Since it is the square of the
tensor force that determines its contribution to the symmetry
energy, tensor forces having larger magnitudes at short distance
affect more significantly the symmetry energy. It is seen that the
variation of the tensor force at short distance affects
significantly the high-density behavior of nuclear symmetry
energy. Including also the kinetic part of the symmetry energy
$\frac{1}{3}\frac{k_F^2}{2m}$, we show in Fig.~\ref{f2} the 
$E_{\rm sym}(\rho)$. The divergent values of the $E_{\rm sym}(\rho)$ are completely due to the
different short-range tensor forces used. To evaluate the accuracy
of the results obtained using Eq.~(\ref{Mac}), we compare in Fig.~\ref{f2} 
predictions from Eq.~(\ref{Mac}) and the Brueckner-Hartree-Fock (BHF)~\cite{Zuo02} both using the Av18 interaction. 
It is seen that essentially over the whole density range considered, the BHF prediction is
about 7 MeV higher. This is qualitatively understandable since the difference in
central forces between the isotriplet $T = 1$ and isosinglet $T =
0$ channels also contribute to the potential part of the symmetry
energy ~\cite{pan72,Gogny}. The comparison here indicates
clearly that indeed, as expected by G.E. Brown and R.
Machleidt~\cite{mac94}, the tensor contribution dominates the
potential part of the nuclear symmetry energy. The 7 MeV
difference can be considered as the systematic error of
predictions based on Eq.~(\ref{Mac}). Thus, it is
clear that the variation of the short-range tensor force leads to
significantly different symmetry energies at supra-saturation densities. 
To be accurate, nevertheless, one should be cautioned that the uncertain short-range tensor 
force is probably not the only reason for the poorly known high-density
behavior of the $E_{\rm sym}(\rho)$. There are also correlations among probably several factors
that may all affect the $E_{\rm sym}(\rho)$ individually. For example, the short-range tensor force also leads to
neutron-proton correlations in SNM~\cite{sch}. Consequently, the single-nucleon momentum distribution obtains a 
high momentum tail that will change the average kinetic energy of nucleons in SNM~\cite{Bethe}, 
and thus the kinetic part of the $E_{\rm sym}(\rho)$~\cite{Xu11}. While this effect is not
considered here, our results based on Eq.~(\ref{Mac}) are interesting and useful for better understanding the role of
tensor forces in determining the $E_{\rm sym}(\rho)$.

In summary, using an approximate expression for the second-order
tensor contribution to the symmetry energy derived earlier by G.E.
Brown and R. Machleidt, we investigated effects of the short-range
tensor interaction on the density-dependence of nuclear symmetry
energy. We found that indeed the tensor force dominates the
potential part of the nuclear symmetry energy. The uncertain
short-range tensor force contributes significantly to the
divergence of the nuclear symmetry energy especially at supra-saturation densities.

We would like to thank L.W. Chen, W. G. Newton, H.-J. Schulze, C. Xu
and W. Zuo for valuable discussions. This work is supported in
part by the US National Science Foundation under grant PHY-0757839
and PHY-1068022, the US National Aeronautics and Space
Administration under grant NNX11AC41G issued through the Science
Mission Directorate, the National Basic Research Program of China
under Grant 2009CB824800, and the National Natural Science
Foundation of China under Grant 10905048.


\end{document}